\documentstyle{article}
\input boxedeps.tex
\SetRokickiEPSFSpecial  %% dvips by Tom Rokicki
\HideDisplacementBoxes

\def\NP{{\it Nucl. Phys.\ }}

\def\PL{{\it Phys. Lett.\ }}
\def\PR{{\it Phys. Rev.\ }}

\def\a{\alpha}

\def\be{\begin{equation}}
\def\eq{\end{equation}}

\def\q2{{\rm QCD}_{2}}
\def\q4{{\rm QCD}_{4}}

\begin{document}
\begin{flushright}
SLAC-PUB-7472\\
CERN-TH/97-88\\
%hep-ph/9705413
\end{flushright}
\vspace{20mm}
\begin{center}
{\LARGE {}\footnote{\baselineskip=13pt
Work partially supported by the Department of Energy, contract
DE--AC03--76SF00515.}Light-Cone Wavefunctions at Small $x$\\}
\vspace{20mm}
{\bf F. Antonuccio${}^*$, S.J. Brodsky${}^{**}$,
S. Dalley${}^{***,}$\footnote{On
leave from: Department of Applied Mathematics and Theoretical Physics,
Cambridge University,
Silver Street, Cambridge CB3 9EW, England.}\\}
\vspace{10mm}
{\em ${}^*$Max-Planck-Institut f\"{u}r Kernphysik, 69029 Heidelberg, Germany
\\ and \\
Institute for Theoretical Physics, University of Heidelberg,
69120 Heidelberg, Germany
\\
\vspace{5mm}
${}^{**}$ Stanford Linear Accelerator Center, \\
Stanford University, Stanford, California 94309 \\
\vspace{5mm}
${}^{***}$Theory Division, CERN, CH-1211 Geneva 23, Switzerland}
\end{center}
\vspace{20mm}
\begin{abstract}
There exist an infinite number of exact small momentum fraction-$x$
boundary conditions on light-cone wavefunctions
of bound states in gauge theory. They are necessary for finite
expectation values of the invariant mass operator and relate components
of the wavefunction from different Fock sectors.
We illustrate their consequences by analyzing the small-$x$ quark
Regge behavior of a heavy large-$N$ meson, finding power-law rise of
unpolarized distributions. The polarized distribution changes sign and
then vanishes with minus the unpolarized Regge intercept.
\end{abstract}
\newpage

\baselineskip .25in

\section{Light-Cone Dynamics}

Light-cone quantization of QCD is a promising tool to describe the
wealth of experimental information about hadronic structure in terms
of quark and gluon degrees of freedom \cite{stan}.
It has
the advantage of dealing explicitly with the hadronic wavefunction
in a general Lorentz frame,  and it is particularly convenient for
analyzing matrix elements of currents such as form factors and light-cone
dominated inclusive processes.
In practice the calculation of a hadronic
wavefunction presents a formidable many-body problem since arbitrary numbers
of gluons and sea quarks can play a significant role.

Large numbers of partons necessarily have large free energy in the
light-cone formalism.
In this paper we shall
point out  some rather simple but important exact restrictions on
the light-cone wavefunctions which follow from high light-cone
energy boundary conditions alone. These `ladder relations' relate
Fock space sectors containing different numbers of partons.
Thus the different Fock components of a hadronic wavefunction in QCD are not
analytically independent.
We give a simple application to the quark distribution function of a heavy
meson as an illustration of the physical consequences.

Each hadronic bound state in the light-cone Hamiltonian formalism of QCD is
an eigenstate $|\Psi(P^+,{\bf P}^{\perp})>$ of the invariant mass
operator $\hat{M}^2 = 2P^+ P^- - |{\bf P}^{\perp}|^2$
where $P^- =
(P^0-P^3)/\sqrt{2}$ is the light-cone energy,  which is the displacement
operator in light-cone time $x^+ =(x^0 + x^3)/\sqrt{2}$,
while $P^+= (P^0 + P^3)/\sqrt{2}$ and ${\bf P}^{\perp}$ are
the conserved total momenta. The operator $P^-$ contains both the kinetic
energy and interaction parts of the light-cone Hamiltonian.  The eigenfunction
of the bound state can be expanded on the Fock basis of free quark and gluons.
The light-cone energy $k^-$ of each such constituent of
mass $m$ carrying light-cone longitudinal momentum $k^+$
and transverse momentum ${\bf k}^{\perp} = (k^1,k^2)$ is
\be
k^{-} = {m^2 + |{\bf k}^{\perp}|^2 \over 2k^+} \ . \label{free}
\eq
The light-cone wavefunction for the $n-$parton state can be labeled by its
constituents' momenta ${\bf k}_{i} = (k^{+}_{i}, {\bf k}^{\perp}_{i})$
and helicities $\a_i$ :
$f_{\a_1 \cdots \a_n} ({\bf k}_1, \cdots , {\bf k}_n)$
where $\sum_i k^{+}_{i} = P^+ $ and $\sum {\bf k}^{\perp}_i = {\bf 0}^{\perp}$.
At first sight,  one would  expect that the finiteness of
the kinetic part of the operator $P^-$ would always force the
light-cone wavefunction for each Fock component to
vanish at $x_i = k^{+}_{i} /P^+ = 0$
for each fermion constituent since $m^2 > 0.$
In fact, we shall show that in theories with Yukawa-like (i.e.
fermion-boson-fermion) interactions such as gauge
theories, that this is not the case.

For example, consider the light-cone
$SU(N)$ gauge theory quantized on the
surface $x^+ = 0$ in the light-cone gauge
$A^{+} = A_- = 0$ with one quark flavor.
It is well-known that $A_+$ and half the
components of the quark spinor, the left-moving half $(v_+,v_-)$ of a
chiral representation say, with $\pm$ chiralities,
are constrained fields. They may be eliminated by their equations of
motion
\begin{eqnarray}
 {\rm i} \partial_{-} v_{\pm} & = & F_{\mp} \label{const1} \\
\partial_{-}^2 A_+ & = &  J  \label{const2}
\end{eqnarray}
where
\begin{eqnarray}
       F_+ & = & {\rm i}(\partial_{z} + {\rm i} g A_{z})u_{-} +
                  \frac{m}{\sqrt{2}} u_{+} \\
       F_- & = & -{\rm i}(\partial_{{\bar z}} + {\rm i} g
                     A_{{\bar z}})u_{+} +
                  \frac{m}{\sqrt{2}} u_{-} \\
       J & = & \partial_-(\partial_{z}A_{{\bar z}} +
            \partial_{{\bar z}} A_{z}) +
              g({\rm i}[A_{z}, \partial_- A_{{\bar z}} ]
               +{\rm i}[A_{{\bar z}}, \partial_- A_{z}]) \nonumber \\
& & \hspace{20mm}
         + \hspace{1mm} g(u_{+}u_{+}^{\dagger} + u_{-}u_{-}^{\dagger})
                  \ .
\end{eqnarray}
In the above expressions
\begin{equation}
     A_z \equiv
 \frac{1}{\sqrt{2}}(A_1 - {\rm i} A_2) \hspace{7mm}, \hspace{7mm}
     \partial_z \equiv
 \frac{1}{\sqrt{2}}(\partial_1 - {\rm i} \partial_2)
     \ ,
\end{equation}
\begin{equation}
     A_{\bar{z}} \equiv
 \frac{1}{\sqrt{2}}(A_1 + {\rm i} A_2) \hspace{7mm}, \hspace{7mm}
     \partial_{\bar{z}} \equiv
 \frac{1}{\sqrt{2}}(\partial_1 + {\rm i} \partial_2)
     \ ,
\end{equation}
and $u_{\pm}$ form respectively the positive and
negative chiralities of the remaining right-moving components of the
quark.
The exchange of non-propagating particles associated with the
constrained fields results in non-local interactions
in the light-cone Hamiltonian
\begin{equation}
P^- =  \int d x^- d{\bf x}^{\perp} \left\{
     F_+^{\dagger} \frac{1}{{\rm i}\partial_-} F_+ +
     F_-^{\dagger} \frac{1}{{\rm i}\partial_-} F_-
  +\frac{1}{2}\mbox{Tr} \left[ - J \frac{1}{\partial_-^2} J
      +    ({\cal F}_{12})^2 \right] \right\} \label{ham}
\end{equation}
where ${\cal F}_{12} = \partial_1 A_2 - \partial_2 A_1 + {\rm i} g [A_1,A_2]$.
We see that the free-fermion kinetic term is replaced by a gauge-field
dependent expression,
\be
 {m^2 \over 2}u^{\dagger} \frac{1}{{\rm i}\partial_-} u \to
F^{\dagger} \frac{1}{{\rm i}\partial_-} F  \ ,
\eq
in analogy with the replacement
\be
{p^2 \over 2m} \to {(p-eA)^2 \over 2m}
\eq
of non-relativistic electrodynamics. Thus the combination
$F$ plays a special role as a fermionic `mechanical velocity' in the gauge
extension of the free kinetic energy of the gauge theory.  The zero momentum
limit of the constraint equation (\ref{const1}) now forces an
interaction-dependent condition on the quark-gluon combined system
\be
\int_{-\infty}^{+\infty} dx^- F_{\pm} = 0 \label{zero} \ ,
\eq
implying that
the field $v$ vanishes at $x^{-} = \pm \infty$. This condition
is necessary for
finiteness of the interactions non-local in $x^-$
involving $F$ in (\ref{ham}),
at fixed transverse co-ordinates
and simply translates into a
condition on the fields at vanishing longitudinal
momentum $k^+ = 0$. However, the finiteness condition does not imply the
individual fixed particle number light-cone wavefunctions have to vanish at
$x_i=0$, but rather that combinations of the wavefunctions involving one
more and one less gluon quanta are related at the small $x$ boundary.

\section{Ladder Relations.}

In order to elucidate the consequences of (\ref{zero}) in the quantum theory,
we can consider its effect
on specific light-cone wavefunctions. Introducing the harmonic
oscillator modes
of the physical fields\footnote{$i,j =1,\ldots,N$ are gauge indices
and $\dagger$ is here understood as the quantum complex conjugate, so
it does not transpose them.}
\begin{eqnarray}
\lefteqn{{A_{z}}_{ij}(x^-,{\bf x}^{\perp})  =
\frac{1}{(2\pi)^{3/2}}\int_0^{\infty} \frac{dk^+}{\sqrt{2 k^+}}
\int d{\bf k}^{\perp}  \times } \nonumber \\
& & \left[
 a_{+ij}(k^+,{\bf k}^{\perp})e^{-{\rm i}(k^+ x^- - {\bf k}^{\perp} \cdot
 {\bf x}^{\perp})} +
 a_{-ji}^{\dagger}(k^+,{\bf k}^{\perp})e^{+{\rm i}
(k^+ x^- - {\bf k}^{\perp} \cdot
 {\bf x}^{\perp})} \right] \\
\lefteqn{{u_{\pm}}_{i}(x^-,{\bf x}^{\perp})  =
\frac{1}{(2\pi)^{3/2}}\int_0^{\infty} dk^+
\int d{\bf k}^{\perp}  \times } \nonumber \\
& & \left[
 b_{\pm i}(k^+,{\bf k}^{\perp})e^{-{\rm i}(k^+ x^- - {\bf k}^{\perp} \cdot
 {\bf x}^{\perp})} +
 d_{\mp i}^{\dagger}(k^+,{\bf k}^{\perp})e^{+{\rm i}
(k^+ x^- - {\bf k}^{\perp} \cdot
 {\bf x}^{\perp})} \right] \ ,
\end{eqnarray}
we can expand any hadron state $|\Psi(P^+,{\bf P}^{\perp})>$
in terms of a Fock basis.
The operators $a_{\pm}^{\dagger}$ create gluons with helicity $\pm 1$,
while $b_{\pm}^{\dagger}$ and $d_{\pm}^{\dagger}$
correspond to quarks and antiquarks
(respectively) with helicities $\pm \frac{1}{2}$.
For concreteness we will consider a
meson in the frame ${\bf P}^{\perp}=0$ in the large $N$ limit,
using a basis of Fock states singlet under residual global gauge
transformations.
At large $N$ there is gluon but not quark pair production, so
a meson is  the superposition of $\bar{q}q$, $\bar{q}gq$,
$\bar{q}ggq$, $\bar{q}gggq$, and so on. Explicitly,
\begin{eqnarray}
\lefteqn{|\Psi(P^+)> =
 \sum_{n=2}^{\infty} \int_0^{P^+} dk_1^+ \dots dk_n^+
\sum_{\alpha_i = \pm}
   \hspace{1mm} \delta(k_1^+ + \cdots + k_n^+ - P^+) \times }
\nonumber \\
& &
\int d{\bf k}^{\perp}_{1} \dots d{\bf k}^{\perp}_{n}
 \hspace{1mm} \delta({\bf k}^{\perp}_{1} + \cdots + {\bf
 k}^{\perp}_{n}  )
\hspace{1mm}
 f_{\alpha_1 \dots \alpha_n}
({\bf k}_1, {\bf k}_2, \dots, {\bf k}_n) \times \nonumber \\
& &  \frac{1}{\sqrt{N^{n-1}}}
 d^{\dagger}_{\alpha_1 i}({\bf k}_1) a^{\dagger}_{\alpha_2 ij}
({\bf k}_2) a^{\dagger}_{\alpha_3 jk}({\bf k}_3) \dots
a^{\dagger}_{\alpha_{n-1} lm}({\bf k}_{n-1})
b^{\dagger}_{\alpha_n m}({\bf k}_n) |0>
\label{mesonbs}
\end{eqnarray}
where repeated indices are summed over and the 
coefficients $f$, depending on the momenta and helicities,
diagonalize $\hat{M}^2$. If one writes the Fock wavefunctions
$f_{\alpha_1 \dots
\alpha_n}$ in terms of the light-cone momentum fractions $x_i=k^+_i/P^+$ and
relative transverse momentum ${\bf k}^{\perp}_i - x_i {\bf P}^{\perp}$ then the
Fock representation is independent of the total momentum $P^+$ and ${\bf
P}^{\perp}$.

\begin{figure}
\centering
\BoxedEPSF{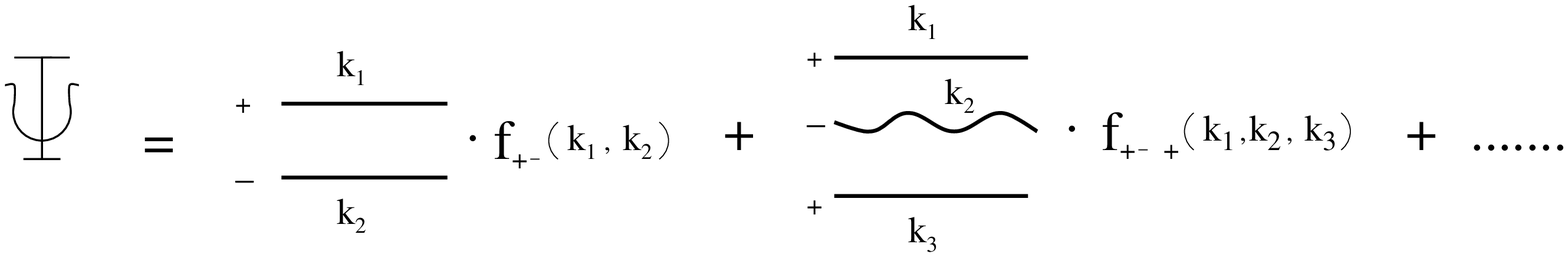 scaled 600}
\caption{Expansion of the large-$N$ meson light-cone wavefunction (for a total
helicity zero example) in terms of parton
Fock components, represented by propagators for quark and anti-quark
(solid) and gluons (wavy).\label{fig1}}
\end{figure}

The quantum version of the statements in the preceding section turns
out to be a little delicate due to operator ordering ambiguities.
We have chosen a prescription which is at least consistent at large
$N$. Introducing
\begin{equation}
  {\tilde F}_{\pm}(k^+,{\bf k}^{\perp}) =
\frac{1}{(2\pi)^{3/2}} \int_{-\infty}^{\infty} dx^- d{\bf x}^{\perp}
\hspace{1mm}   {F}_{\pm}(x^-,{\bf x}^{\perp})
e^{-{\rm i}(k^+ x^- - {\bf k}^{\perp} \cdot
 {\bf x}^{\perp})}
\end{equation}
we find that (\ref{zero}) can be meaningfully applied as an annihilator of
physical states for the cases
\begin{eqnarray}
 \lim_{k^+ \rightarrow 0^-}    {\tilde F}_{\pm i}
 (k^+, {\bf k}^{\perp}) \cdot |\Psi(P^+,
      {\bf P}^{\perp})> & = & 0 \label{quark} \\
 \lim_{k^+ \rightarrow 0^+} {\tilde F}_{\pm i}^{\dagger}
(k^+, {\bf k}^{\perp}) \cdot |\Psi(P^+,
      {\bf P}^{\perp})> & = & 0 \label{aquark} \ .
\end{eqnarray}
The first relation yields a condition on the Fock space wavefunctions $f$
involving vanishing quark longitudinal momentum, the second
on vanishing anti-quark
longitudinal momentum:
\begin{eqnarray}
\lefteqn{m f_{\mp \pm \alpha_1 \cdots \alpha_n}
({\bf k},{\bf k}_1,\dots,
{\bf k}_{n+1})} & & \nonumber  \\ \lefteqn{\pm
(k^1 \pm {\rm i}k^2)f_{\pm \pm \alpha_1 \cdots \alpha_n}
({\bf k},{\bf k}_1,\dots,
{\bf k}_{n+1})
} & & \nonumber \\
& = & \pm \frac{g \sqrt{N}}{(2 \pi)^{3/2}} \left[
 \frac{f_{\pm \alpha_1 \cdots \alpha_n}
  ({\bf k} + {\bf k}_1, {\bf k}_2, \dots, {\bf k}_{n+1})}
    {\sqrt{k_1^+}} \right. \nonumber \\
& & +
\int_0^{\infty}\frac{dp^+ dq^+}{\sqrt{q^+}}
\hspace{1mm} \delta (p^+ + q^+ - k^+) \int d{\bf p}^{\perp}
 d{\bf q}^{\perp} \hspace{1mm} \delta ( {\bf p}^{\perp}
 + {\bf q}^{\perp} - {\bf k}^{\perp}) \times \nonumber \\
& & \hspace{8mm} f_{\pm \mp \pm\alpha_1 \cdots \alpha_n}
({\bf p}, {\bf q}, {\bf k}_1, \dots , {\bf k}_{n+1})
\left.  \frac{}{}  \right] \label{first}
\end{eqnarray}
and
\begin{eqnarray}
\lefteqn{m f_{\mp \mp \alpha_1 \cdots \alpha_n}
({\bf k},{\bf k}_1,\dots,
{\bf k}_{n+1})} & & \nonumber  \\ \lefteqn{\pm
(k^1 \pm {\rm i}k^2)f_{\pm \mp \alpha_1 \cdots \alpha_n}
({\bf k},{\bf k}_1,\dots,
{\bf k}_{n+1})
} & & \nonumber \\
& = & \pm \frac{g \sqrt{N}}{(2 \pi)^{3/2}}
\int_0^{\infty}\frac{dp^+ dq^+}{\sqrt{q^+}}
\hspace{1mm} \delta (p^+ + q^+ - k^+) \int d{\bf p}^{\perp}
 d{\bf q}^{\perp} \hspace{1mm} \delta ( {\bf p}^{\perp}
 + {\bf q}^{\perp} - {\bf k}^{\perp}) \times \nonumber \\
& & \hspace{8mm} f_{\pm \mp \mp \alpha_1 \cdots \alpha_n}
({\bf p}, {\bf q}, {\bf k}_1, \dots , {\bf k}_{n+1}) \ ,
\label{second}
\end{eqnarray}
with a similar set of
relations for quarks;  in (\ref{first}) and (\ref{second})
${\bf k} = (k^+,{\bf k}^{\perp})$ and the limit
$k^+ \to 0^+$ is understood. We can interpret the above relations as
the leading result for small but finite $k^+$, with  corrections
at higher order in $k^+/k^{+}_{1}$ etc.
Similar `ladder relations' were first
considered in the context of a `collinear' one-space, one-time approximation to
light-cone QCD
\cite{coll,houch}
 If we adopt the following momentum-space
operator ordering in $P^-$
(\ref{ham})
\be
\int d{\bf k}_{\perp} \left\{ - \int_{-\infty}^{0} {dk^+ \over k^+}
  \left( {\tilde F}_{+ i}^{\dagger} {\tilde F}_{+ i}
 +  {\tilde F}_{- i}^{\dagger} {\tilde F}_{- i} \right)
  +
\int_{0}^{\infty} {dk^+ \over k^+}
 \left( {\tilde F}_{+ i} {\tilde F}_{+ i}^{\dagger}
+ {\tilde F}_{- i} {\tilde F}_{- i}^{\dagger} \right)
\right\} \ , \label{form}
\eq
this manifestly ensures finiteness at the $k^+ = 0$ pole.
%Conversely, starting from (\ref{form}),
%if we demand that $<\Psi| P^- | \Psi>$ is finite, then
%(\ref{quark}) and (\ref{aquark})
%follow, since our Fock space has only positive norm states.
Normal ordering
the oscillator modes in  $P^-$ would spoil finiteness. Since we do not use
normal ordering of the form (\ref{form}),
infinite quark self energies (self-inertias)
are generated but no vacuum energies
are generated.
One sees explicitly from (\ref{first}) and (\ref{second})
that the wavefunction components with at least one
gluon do not in general vanish for small quark or antiquark $k^+$.
This  is an intrinsic
property of the bound state, i.e. no reference has been made to
perturbation theory or special ${\bf k}_{\perp}$ kinematic regimes. It is
a boundary condition on wavefunctions necessary for finite expectation value
of the invariant mass operator.

The wavefunction components $f$ are the
solutions to the bound state problem as represented by a light-cone
relativistic Schr\"{o}dinger many-body matrix equation: if ${\cal M}$
is the bound state mass eigenvalue, projecting $\hat{M}^2 |\Psi> = {\cal
M}^2 |\Psi>$  onto a specific $n$-parton Fock state one derives
\be
\left({\cal M}^2
- \sum_{i=1}^{n} \left[{({\bf k}_i^{\perp})^2 + m_i^2 \over
 x_i}
\right]\right)  f_{\alpha_1 \dots \alpha_n}
({\bf k}_1, {\bf k}_2, \dots, {\bf k}_n) = \hat{V}\left[
f_{\alpha_1 \dots \alpha_n}
({\bf k}_1, {\bf k}_2, \dots, {\bf k}_n)\right]
\label{full}
\eq
with interaction kernel $\hat{V}$ (including self-inertias).
In eq.(\ref{full}) a small momentum limit $x_i = k^+_i/P^+  \to 0$
for given $i$
also has the potential to yield a boundary condition
on the wavefunction independent of the mass ${\cal M}$. However this
is equivalent to imposing that $\sim   \tilde{F}^{\dagger} \tilde{F} $
at zero momentum should annihilate
physical states, evidently a more complicated
condition than eq.(\ref{quark}).
Eq.(\ref{full}) does have the advantage that one can consider
more than one
parton having small momentum and corrections to the small momentum limit.

So far, the parameters in $\hat{V}$ have been interpreted as the bare
ones of an ultraviolet-regulated Hamiltonian.
When the quark small-$x$ boundary approaches another high-energy corner
of phase space, such as large transverse momentum or gluon small $x$, further
counterterms must be added to obtain finite answers.
The
wavefunctions which are integrated in the relations
(\ref{first}) and (\ref{second}) involve {\em two} vanishingly small
longitudinal
momenta as $k^+ \rightarrow 0^+$.  Because of this,
although the integration domain has vanishingly small measure,
the integrands are sufficiently singular to give a non-zero result.
To determine the
singular behavior of the wavefunction components appearing in the
integrand requires us in general to study the renormalization
of the light-cone Schr\"{o}dinger equation (\ref{full}), taking  account
of all the high energy cut-offs which must be applied, which is beyond
the scope of this paper. However
the net effect will be virtual corrections
to non-integral terms in (\ref{first}) and (\ref{second}).
One uses (\ref{full}) to express the integrands in terms of
different Fock sector components. This generates
corrections to the non-integral terms in
(\ref{first}) and (\ref{second}) plus further
integrals involving more partons and/or higher powers of the coupling.
The above procedure may then be repeated iteratively for the latter,
so generating renormalizations of the non-integral terms
in (\ref{first})(\ref{second}) in powers of the coupling
constant; additional non-integral terms are also generated
with more exotic helicity structure.
This procedure is demonstrated with examples in the context of the
collinear approximation to QCD in ref.\cite{houch}.

For the remainder of this paper we shall consider a `tree-level' approximation
to the ladder relations, dropping the integral terms. This should be a good
approximation for heavy quarks, when the expansion parameter for loop
corrections is $g/m$. In this regime we may also neglect helicity
non-conserving pieces, i.e. we consider transverse momenta bounded
by $\Lambda^{\perp}$ where $\Lambda^{\perp}/m <<1$.
The new ladder relations in the heavy quark regime may then be written
\begin{eqnarray}
f_{\pm \pm \alpha_1 \cdots \alpha_n}
({\bf k},{\bf k}_1,\dots,
{\bf k}_{n+1}) & = & 0
\label{rel3} \\
f_{\mp \pm \alpha_1 \cdots \alpha_n}
({\bf k},{\bf k}_1,\dots,
{\bf k}_{n+1}) & = &
 \pm \frac{g \sqrt{N}}{m(2 \pi)^{3/2}}  \left[
 \frac{f_{\pm \alpha_1 \cdots \alpha_n}
  ({\bf k} + {\bf k}_1, {\bf k}_2, \dots, {\bf k}_{n+1})}
    {\sqrt{k_1^+}} \right]
\label{ladren}
\end{eqnarray}
where ${\bf k} = (k^+ \to 0^+,{\bf k}^{\perp})$ as before.

\section{Small-$x$ Distribution Functions.}

We now shall show that the ladder relations reproduce
results for the leading $\log{1/x}$  approximation of quark and
anti-quark distribution
functions, which are usually obtained by summing ladder diagrams \cite{glr}.
For illustration, we shall calculate these distributions
for  heavy-quarkonium (\ref{mesonbs}) in the large-$N$ limit. 
Following ref.\cite{lepage} we
define the probability of finding an anti-quark
with longitudinal momentum fraction $x=k^+/P^+$ and
collinear with the hadron up to scale $\Lambda^{\perp}$ as
\begin{eqnarray}
Q(x, \Lambda^{\perp})  & = &  \sum_{n=2}^{\infty} \sum_{\alpha_i}
 \int_0^{P^+} dk_1^+ \dots dk_n^+
   \hspace{1mm} \delta(k_1^+ + \cdots + k_n^+ - P^+) \times \nonumber \\
& &
\int_{0}^{\Lambda^{\perp}} d{\bf k}^{\perp}_{1} \dots d{\bf k}^{\perp}_{n}
 \hspace{1mm} \delta({\bf k}^{\perp}_{1} + \cdots + {\bf
 k}^{\perp}_{n} )
\times \nonumber \\
& &  \delta (k_1^+ - k^+) \hspace{1mm}
 |f_{\alpha_1 \dots \alpha_n}
({\bf k}_1, {\bf k}_2, \dots, {\bf k}_n)|^2
\label{qxsf}
\end{eqnarray}
(the analysis may be repeated trivially for quarks).
In this {\em intrinsic} contribution to the distribution function,
$\Lambda^{\perp}$ is not necessarily large,  but it could be used as the input
for an {\em extrinsic} evolution to large momentum scales.

Consider the contribution to $Q(x \to 0)$ in a helicity $+1$ polarized
meson from  partons with alternating
helicity $f_{++}$, $f_{-++}$, $f_{+-++}$, $f_{-+-++}$, $\ldots$.  
For heavy quarks only these components of the wavefunction will contribute as
a result of eqs.(\ref{rel3})(\ref{ladren}) and the dominance of $f_{++}$ in the
valence part if we assume zero orbital angular momentum $L=0$.
If $k^+ << k_2^+$ in (\ref{qxsf}) we may use the ladder relation
(\ref{ladren}) to re-express the $n$-parton wavefunction in terms of
that for $n-1$ partons. Evidently, the new integrand in (\ref{qxsf}) gives
its dominant contribution in the region of small $k_2^+$, so we may apply the
ladder relation again. This process may be iterated until one arrives
at the $\bar{q}q$ valence wavefunction. Consider therefore the
contribution from the integration region
\be
\int_{{\cal C}^+} =  \int_{k_1^+} dk_2^+ \int_{k_2^+} dk_3^+
\cdots \int_{k_{n-3}^+} dk_{n-2}^+  \ .
\eq
Then the $n$-parton contribution to (\ref{qxsf}) is approximately
\be
\left(\frac{g^2 N}{8 \pi^3} \right)^{n-2}
\int_{{\cal C}^+} \int_{{\cal C}^{\perp} }
{I_{n-2}  \over k_2^+ k_3^+ \cdots k_{n-2}^+}
 \int_0^{P^+} dk_n^+ \int d{\bf k}^{\perp}_n
\frac{|f_{++}((P^+ - k_n^+, -{\bf k}_n^{\perp}),
                    {\bf k}_n)|^2}{P^+ -  k_n^+}
\eq
where
\be
\int_{{\cal C}^{\perp}} I_{n-2} =
\int \frac{ d{\bf k}^{\perp}_1}{m^{2}} \cdot \int \frac{d{\bf k}^{\perp}_2 }
{m^{2}} \cdot  \cdots \cdot  \int
\frac{d{\bf k}^{\perp}_{n-2} }{m^{2}}
\eq
For simplicity we have used the same (global)
transverse cut-off $|{\bf k}^{\perp}| \leq \Lambda^{\perp}$ for each parton.
The heavy quark limit is a choice rigorously compatible with the leading
$\log{1/x}$ approximation, namely
$(g^2)^{n-1} \int I_{n-1} << (g^{2})^n \int I_n$.  
\footnote{If $\Lambda^{\perp} >> m$, then
it is the wavefunctions with totally aligned helicities which
dominate in the distribution  functions ---  we would need the most general
form of the ladder relations now ---
but we cannot be sure that taking leading
$\log{1/x}$'s is accurate in this case because the momentum dependence
of the corresponding transverse
integrals $I_n$ would tend to cancel
the asymptotic freedom of the running coupling.}

Thus
\begin{eqnarray}
Q(x \to 0 ) & \approx & \frac{g^2 N}{8 \pi^3} \cdot <P^+/k^+_{n}>_{\rm val}
\sum_{n=3}^{\infty}
\frac{1}{(n-3)!}\left[\frac{g^2 N}{8 \pi^3}
 \ln \frac{1}{x} \right]^{n-3} \int_{{\cal C}^{\perp}} I_{n-2}
\label{isum} \\
 & = &  <P^+/k^+_{n}>_{\rm val}
\frac{g^{2} N (\Lambda^{\perp})^2}{8 m^{2} \pi^2}
x^{-g^{2} N (\Lambda^{\perp})^2/8 m^{2} \pi^2} \label{ans}
\end{eqnarray}
where $<P^+/k^+_{n}>_{\rm val}$ is the expectation value
of the inverse anti-quark momentum
fraction in the $\bar{q}q$ valence sector. Note that this result
ignores the contribution
to $Q$ from the valence wavefunction itself, which is expected to vanish
at small $x$.
The answer (\ref{ans}) is similar  to that of
Reggeon exchange between virtual photon and
hadron which one could have obtained by summing appropriate ladder
graphs under the same assumptions. 
We would have to generalize our
ladder relation to finite $N$ and $m$, thus allowing sea quark pair production,
in order to observe pomeron-like contributions in the quark structure
function of the meson; alternatively such contributions could be seen in
the gluon structure function of a heavy large-$N$ meson,
as has been shown by Mueller \cite{ahm}.

The corresponding
polarized  (anti-)quark distribution function $\Delta Q$ has an extra
${\rm sgn}(\a_1)$ factor in the definition (\ref{qxsf}), leading
to an extra  $(-1)^n$ in the
sum (\ref{isum}).
Thus the polarization asymmetry for a helicity $+1$
meson in the leading $\log { 1/x}$ heavy quark approximation behaves as
\be
{\Delta Q \over Q}(x \to 0) \approx -x^{g^{2} N (\Lambda^{\perp})^2/4 m^{2}
\pi^2} \label{neg}
\eq
and is negative because the dominant process at small $x$ for heavy
quarks is the helicity-flip emission of one gluon. Note that this implies
that $\Delta Q$ has convergent Regge behaviour with minus the Regge
intercept of $Q$.
This result again neglects the 
direct contribution of the  $\bar{q}q$
sector, which is also expected to vanish at small $x$, only more quickly
than (\ref{neg}).
The convergent behavior (\ref{neg}), although a correct
consequence of our approximation,  may be unrealistic for physical
applications given
that our approximation (large N, heavy quarks) neglects quark pair
production. That, and sub-leading log contributions, could 
upset the delicate cancelations which lead to vanishing $\Delta Q$ above.
We shall assume nevertheless that the ladder relations give the 
correct sign of $\Delta Q(x)$ at small $x$.
Since the helicity of any parton as $x \to 1$ tends to align with that of the
hadron\cite{BBS}, the polarization asymmetry should then change sign at
sufficiently
small $x$ due to (\ref{neg}). This sign change has been
explicitly observed in the non-perturbative solution of the collinear
model for QCD\cite{houch}.

These considerations should also be relevant to the
sign of the heavy sea quark polarization in the nucleon.
For example, strange quarks are predicted to be helicity-aligned with the
nucleon
polarization at $x \to 1$\cite{BBS}. However, most of the strange quark
distribution occurs at small $x$, where according to the ladder relations and
Eq. (\ref{neg}), $\Delta s(x) $ will have a negative sign.  
We also note that explicit models of the
intrinsic strangeness distribution of the nucleon\cite{Sig87,Bro96} predict
a negative
$\Delta s$ for the strange quarks\cite{Bro96}.
Thus we also expect a
change in sign of the strange quark helicity distribution of the
nucleon at an intermediate value of $x$.

\section{Conclusions.}

The methods of this paper allow the derivation of generalized ladder relations
and their corrections for any number of partons by
considering small-$x$ expansions  of the renormalized light-cone bound-state
equation (\ref{full}).
As we have seen, the leading orders of this expansion can yield
interesting relations and information on the Reggeon structure of quark
structure functions which are formally independent of the details of the bound
state, such as its mass
${\cal M}$.  The analysis given in this paper of heavy quarkonium ladder
relations demonstrates the origin of the Regge behavior of QCD for both
the polarized and unpolarized structure functions, although further work will
be required to show that this gives the rigorous behavior at $x \to 0.$
We have also applied the ladder relations to
derive constraints on the Regge behavior and sign of the polarized
distributions at $x \to 0.$ 
These results also have interesting phenomenological
predictions for the polarization correlations of non-valence
quark and anti-quark distributions.

It is also possible to develop further relations between Fock components of the
light-cone wavefunctions which follow from the zero momentum limit of matrix
elements of the current
$J$ in eq.(\ref{const2}). Inspection of Eq.(\ref{ham}) shows that
this is again a
finite energy condition; i.e., it is a boundary condition for small gluon
$k^+$ \cite{ahm}.

We also note that the behavior of light-cone wavefunctions at
very small
$x$ are  generally difficult to obtain
by numerical solution of the discretized
bound-state problem without enormous computational cost. However, the
analytic ladder relations derived here may be usefully employed to extend the
coarse-structured numerical results into otherwise inaccessible regions
of phase
space.

\vfil
\end{document}